\documentclass[prb,twocolumn,showpacs]{revtex4}

\usepackage{psfrag}
\usepackage{times}
\usepackage{amsmath}
\usepackage{amssymb}
\usepackage{graphicx}

\begin{document}

Published in {\em Applied Physics Letters} {\bf 78}, 3316-3318 (2001)

\vspace*{0.5cm}

\title{Production of ordered silicon nanocrystals by low-energy ion 
sputtering}

\author{Ra\'ul Gago and Luis V\'azquez\footnote[1]{Electronic mail: 
lvb@icmm.csic.es}}
\affiliation{Instituto de Ciencia de Materiales de Madrid (CSIC),
Cantoblanco, 28049 Madrid, Spain}
\author{Rodolfo Cuerno}
\affiliation{Departamento de Matem\'aticas \& GISC, Universidad 
Carlos III de Madrid, Avenida Universidad 30, 28911 Legan\'{e}s, Spain}
\author{Mar\'{\i}a Varela and Carmen Ballesteros}
\affiliation{Departamento de F\'{\i}sica, Universidad Carlos III de Madrid, 
Avenida Universidad 30, 28911 Legan\'{e}s, Spain} 
\author{Jos\'e Mar\'{\i}a Albella}  
\affiliation{Instituto de Ciencia de Materiales de Madrid (CSIC),
Cantoblanco, \mbox{} 28049 Madrid, Spain}

\begin{abstract}
\vspace*{0.5cm} We report on the production of ordered assemblies of silicon
nanostructures by means of irradiation of a Si(100) substrate with 1.2 keV
Ar$^+$ ions at normal incidence. Atomic Force and High-Resolution
Transmission Electron microscopies show that the silicon structures are 
crystalline, 
display homogeneous height, and spontaneously arrange into short-range 
hexagonal ordering. 
Under prolonged irradiation (up to 16 hours) all dot characteristics remain 
largely unchanged and a small corrugation develops at long wavelengths. 
We interpret the formation of the dots as a result of an instability due to
the sputtering yield dependence on the local surface curvature.
\end{abstract}



\maketitle

The production of semiconductor nanostructures has attracted the
interest of many research groups because of the important applications
in optoelectronic and quantum devices.\cite{1} Many of the
interesting materials properties depend on the size, shape and
regularity of the  nanometric substructure. In particular, efficient
light emission from silicon is achieved when the nanostructures are
smaller than the bulk exciton.\cite{2,3} Silicon nanostructures are
very important for the development of new types of microelectronic,
electrooptical, electrochemical, electromechanical, sensing 
and silicon laser devices.\cite{3,4,5} Techniques such as Si ion
implantation, e-beam writing, scanning probe litography, pulsed laser
deposition, laser annealing, low pressure chemical vapor deposition
and thermal evaporation have been used to fabricate silicon
nanostructures.\cite{3,6} However, it is not evident that any of 
these techniques 
can be used to simultaneously control purity, uniformity and crystallinity
of the nanostructures and to produce these efficiently in a large
scale.  Thus, new processes of fabrication of silicon nanocrystals are
being investigated.\cite{4,7} In this sense both the ease and
reproducibility of the process involved are key factors for its
eventual practical use. One very promising candidate is irradiation of
a monocrystalline semiconductor surface by low energy ions. This
technique has already proved its capability for the production of
self-organized quantum dots on GaSb,\cite{8} and InP surfaces,
\cite{9} with promising photoluminescence properties.\cite{8} 
However, although the production of ripple structures on
silicon surfaces has been reported,\cite{10} to our knowledge, no
report exists on the application of this irradiation technique to
produce nanocrystalline dots on silicon surfaces.

In this letter we report on the production and morphological
characterization of nanocrystalline silicon dots on Si (100) surfaces
by 1.2 keV Ar$^+$ ion bombardment. We have produced these nanostructures
over a wide temporal range spanning from two minutes up to 16
hours. These experiments have allowed us to confirm the morphological
stability of the features produced, as well as to observe the
development of a long wavelength corrugation. As a result we interpret
the physical mechanism determining the surface morphology to be an
instability due to the local surface curvature dependence  of the
sputtering yield.\cite{11,12}

The samples were processed in a high vacuum chamber with a
base pressure of $2\times 10^{-7}$ mbar. Si (100) wafers were placed on
the sampleholder and irradiated with an Ar$^+$ beam from a 3 cm
Kauffman ion gun. Argon ions, accelerated at 1.2 keV, impinged
normally on the silicon surface with an effective ion current
density of 0.24 mA/cm$^2$. The irradiation time was varied between
1 minute and 16 hours. The bombarded samples were studied by
Atomic Force Microscopy (AFM) and High-Resolution Transmission Electron
Microscopy (HRTEM). AFM measurements were performed in air in
tapping mode, using silicon cantilevers (10 nm radius). 
Specimens suitable for HRTEM were prepared by standard procedures and 
examined using a Philips CM200 FEG analytical microscope operating at 200 kV.
\begin{figure}[ht]
\begin{center}
\includegraphics[width=7cm]{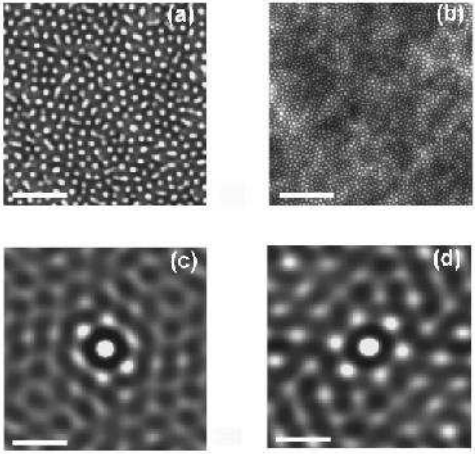}
\caption{AFM images of a Si(100) substrate sputtered by 1.2 keV Ar$^+$ ions at
normal incidence for 6 minutes [(a): $1 \times 1$ $\mu$m$^2$] and
960 minutes [(b): $3 \times 3$ $\mu$m$^2$]. (c) and (d): Two dimensional
autocorrelation functions obtained from $400\times 400$ nm$^2$ areas of images
(a) and (b), respectively. The bars represent 277 nm [panel (a)], 831 nm
[panel (b)] and 111 nm [panels (c) and (d)].}
\label{fig1}
\end{center}
\end{figure}
 
Fig.\ \ref{fig1} shows AFM images of samples bombarded for 6 minutes [panel
(a)] and 960 minutes [panel (b)]. 
Silicon dots, $6 \pm 1$ nm high and 40-50 nm 
of diameter are clearly seen. The AFM images suggest two effects: {\em first}, 
the nearest neighbor distance between dots tends to be a time independent 
constant; {\em second}, the nanostructures self-organize into short-range
hexagonal symmetry, even after 16 hours of sputtering. 
These two facts, which imply a constant dot surface density of 
$\approx 3.5 \times 10^{10}$ dots/cm$^2$, are confirmed in panels (c),
(d) of Fig.\ \ref{fig1}, where the two-dimensional
autocorrelations of the AFM images 
show short-range hexagonal ordering with a nearest dot
distance of $\approx 55$ nm. As seen in Fig.\ \ref{fig1} (b), after 
960 minutes of irradiation, the dot structure is clear but the
surface also displays a long wavelength corrugation, $\approx 500$
nm wide and $\approx 6$ nm high. 

\begin{figure}[th]
\begin{center}
\includegraphics[width=7.0cm]{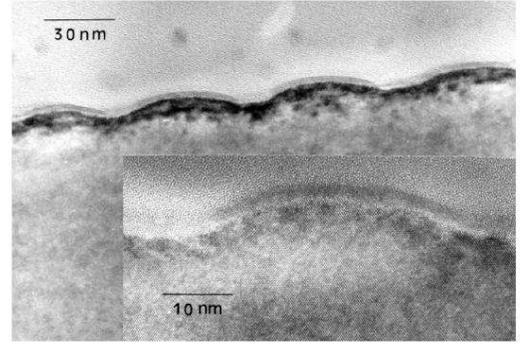}
\caption{Cross-sectional HRTEM multi-beam image along the $\langle 110
\rangle$ direction of a sputtered sample. Ar$^+$ ions at 1.2 keV were
employed at normal incidence for 10 minutes.
Inset: high-resolution image of one of the nanostructures on the
main panel. $\{111\}$ lattice fringes show the crystallinity of the
nanostructures; a low number of defects are visible.}
\label{fig2}
\end{center}
\end{figure}

An important issue concerning the potential application of these dot
structures is their crystallinity.\cite{4,5} 
A cross-sectional multi-beam TEM image of a 10 minutes sputtered
sample along the $\langle 110 \rangle$ direction is shown in Fig.\
\ref{fig2}.  
The formation of uniformly distributed nanostructures
with lenticular shape covered by an amorphous layer $\approx 2$ nm
thick is observed.  The inset of Fig.\ \ref{fig2} shows a
high-resolution image of  one of the nanostructures on the main
panel. \{111\} lattice fringes are visible, showing the crystallinity of the
nanostructures and a low number of defects. The height of
the nanocrystals, that is excluding the amorphous layer, is 
in the range 6-7.5 nm, its width being in the range 40-60 nm.  
These figures are in
agreement with those obtained by AFM provided that tip convolution
effects are considered.  The silicon dots shape and size induced by
ion sputtering  are different from those observed for GaSb,\cite{8}
which are $30 \times 30$ nm conical dots. We note that for the GaSb
system a preferential sputtering  of the Sb atoms was observed, which
is not the case for our system. This fact could be related to the
different dot morphologies observed for both cases.

\begin{figure}
\begin{center}
\includegraphics[width=7.0cm]{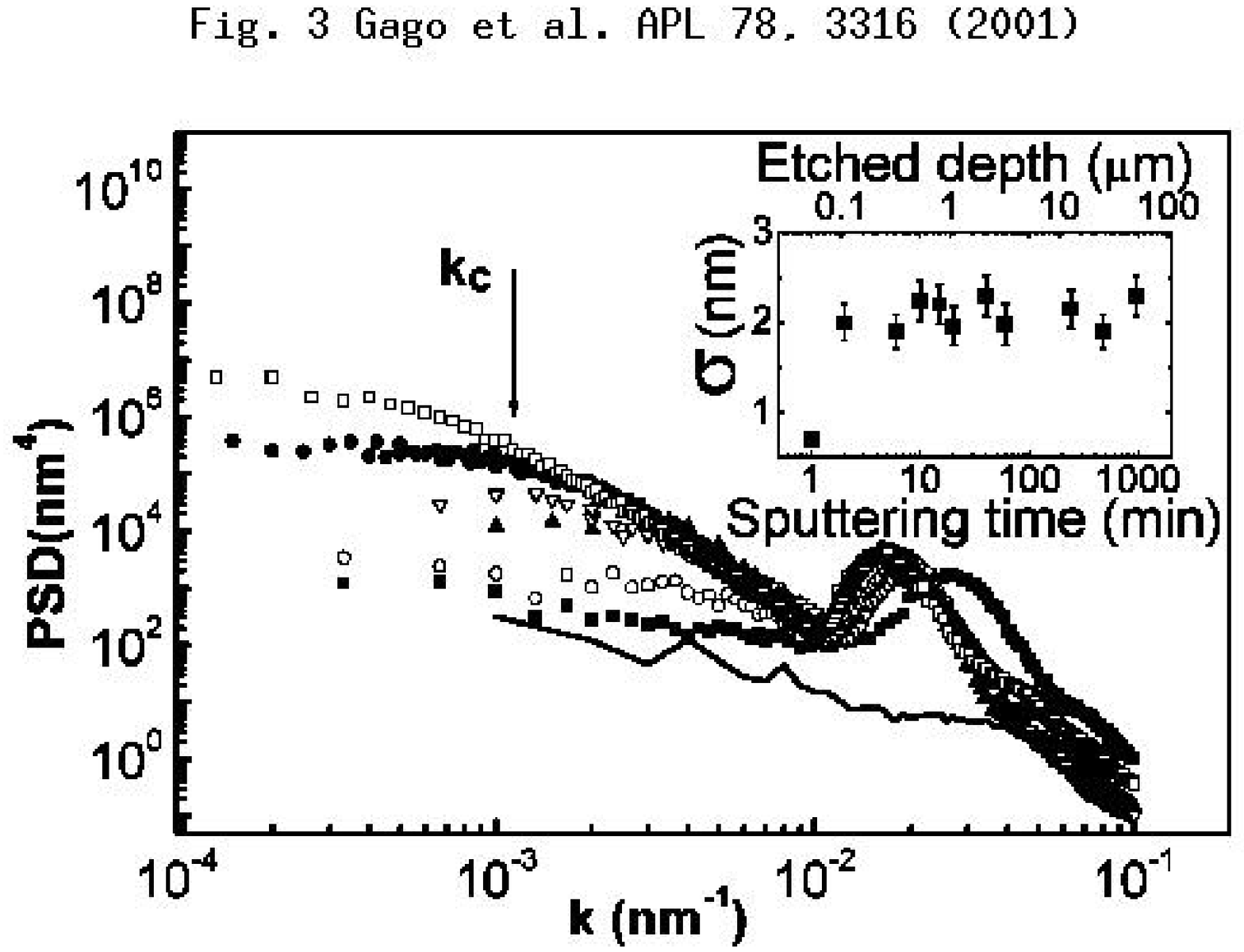}
\end{center}
\caption{Log-log plot ot the PSD curves obtained from AFM images of
Si(100) substrates sputtered by 1.2 keV Ar$^+$ ions at normal incidence
for 1 (solid line), 2 ($\blacksquare$), 6 ($\bigcirc$), 20 ($\blacktriangle$),
60 ($\triangledown$), 240 ($\bullet$), and 960 ($\square$) minutes.
$k_c$ is indicated for the sample irradiated for 960 minutes.
Inset: Surface roughness $\sigma$ versus sputtering time and etched depth.
One minute sputtering time amounts to an ion dose equal to $9 \times 10^{16}$
cm$^{-2}$.}
\label{fig3}
\end{figure}

In order to obtain a better insight into the mechanism of
dot formation for silicon we have prepared samples for times
ranging from 1 minute up to 16 hours. The samples were
systematically analyzed by AFM. The morphological data were
quantitatively analyzed within the framework of the dynamic scaling
theory.\cite{13,14} An important magnitude is the root mean square 
{\em roughness} of the surface, $\sigma$. The dynamic scaling theory 
predicts that a typical lateral length scale $\xi$ of the
system grows as $\xi \sim t^{1/z}$, where $t$ is the
sputtering time and $1/z$ is a coarsening exponent.
As a consequence, when probing the surface over a distance $L > t^{1/z}$, 
the roughness grows as $\sigma \sim t^{\beta}$, whereas for $t^{1/z} > L$ 
the roughness saturates to $\sigma \sim L^{\alpha}$. 
Thus, the exponents $\alpha$ and $z$ describe the interface
evolution. The ratio $\beta \equiv \alpha/z$ can be obtained directly by 
plotting the time evolution of the roughness $\sigma$, as measured by 
AFM; moreover, $\alpha$ can be obtained from the analysis of the power spectral
density (PSD) of the surface morphology measured by AFM, since
the PSD behaves, for a two-dimensional rough surface,\cite{14} as
PSD$(k) \sim  1/k^{2\alpha+2}$, with $k = 1/L$. Fig.\ \ref{fig3} is a 
logarithmic plot of the PSD of the surface morphology measured for 1, 2, 6, 20,
60, 240, and 960 minutes of sputtering. 
The large peak in the
PSDs corresponds to the average distance between nearest neighbor dots. 
We observe that this
wavelength increases with time from $\approx 35$ nm for 2 minutes
until it reaches a value in the 52-60 nm range for times longer than 
6 minutes. For sputtering times longer than 20
minutes the PSD displays for small $k$ values an exponent
value $\alpha = 0.8 \pm 0.1$. For each time there is a minimum wavelength
$k_c$ below which the PSD crosses over to a different behavior. We obtain that
$k_c$ decreases as $k_c \sim t^{-0.2}$. Hence, length scales larger than the 
nearest dot distance and smaller than $1/k_c$ display scale invariant behavior
with exponents $\alpha \simeq 0.8$ and $1/z \simeq 0.2$. This corresponds to 
the development of a long wavelength corrugated structure, which appears after 
20 minutes of sputtering and coarsens with time, see Fig.\ \ref{fig1} (b).
 
Finally, the inset of Fig.\ \ref{fig3} shows the change of $\sigma$ with the
the sputtering time. The roughness
increases sharply during the very first stages of the
sputtering process, but rapidly attains a constant value $\approx
2$ nm. This extremely small roughness is obtained even after a 50
micron thick layer of the silicon wafer has been etched away
by the Ar$^+$ ions. We have found that for long sputtering times
the main morphological difference is the development 
of the long wavelength corrugation. However, the PSD curves 
(Fig.\ \ref{fig3}) indicate that
the main contribution to the total surface roughness $\sigma$ is due to
the dot nanostructures, rather than to the long wavelength corrugation.

The observed behavior is in qualitative agreement with the picture of the 
sputtering process based on the interplay between roughening induced by ion 
sputtering and smoothening due to surface diffusion.\cite{11,12,15,16}  
The instability due to the local surface curvature dependence of the 
sputtering yield \cite{11} produces at short times a regular array of dots 
(with defects induced by shot noise in the ion beam \cite{12}) and a rapid  
increase of the roughness. A similar qualitative behavior has been reported  
for the GaSb and InP dots.\cite{8,9} At longer times, there are nonlinear 
effects \cite{12}
inducing drastically slower growth for the roughness \cite{15,16} as
well as scaling behavior at length scales larger than the nearest dot
distance, of the type encountered for the long wavelength corrugation
[Fig.\ \ref{fig1} (b)]. 
In Ref.\ \onlinecite{16}, 
the production of either dots or holes by ion-sputtering
has been predicted to depend on the
ion penetration depth and the shape of the collision cascade. 
Our experimental results agree quite well with the overall scenario 
thus predicted for the case of a positive nonlinearity. It will be 
interesting to test \cite{17} further 
predictions \cite{16} on morphology behavior as a function of
temperature and ion energy.

In summary, we have shown that crystalline dots ($\approx 7$ nm high
and $\approx 50$ nm wide) are produced on a Si(100) surface by low
energy Ar$^+$ ion bombardment at normal incidence. The silicon
nanocrystals arrange themselves into hexagonal short-range ordering
from 2 minutes up to 960 minutes of sputtering. The main effect of
prolonged sputtering is the development of a long wavelength
corrugation displaying scaling  properties, which does not largely
contribute to the total surface roughness. The experimental
observations as well as the dynamic scaling analysis of the AFM images
are consistent with the formation of the dots as a result of an
instability due to the curvature dependence of the sputtering yield.

This work has been partially supported by DGES (Spain) grants
MAT1999-0830-C03-01, MAT2000-0375-C02-02, and BFM2000-0006.


\begin{thebibliography}{99}

\bibitem{1} G. Schmidt and L.F. Chi, Advanced Materials {\bf 10}, 515 (1998).

\bibitem{2} E. Chomski and G.A. Ozin, Advanced Materials {\bf 12}, 1071 (2000).

\bibitem{3} L. Pavesi, L. Dal Negro, C. Mazzoleni, G. Franzò and
F. Priolo, Nature {\bf 408},  440 (2000).

\bibitem{4} Y. Yin, B. Gates and Y. Xia, Advanced Materials {\bf 12}, 
1426 (2000). 

\bibitem{5} G.F. Grom, D. J. Lockwood, J.P. McCaffrey, H.J. Labb\'e,
P.M. Fauchet, B. White Jr, J. Diener, D. Kovalev, F. Koch, and
L. Tsybeskov, Nature {\bf 407}, 358 (2000).

\bibitem{6} S. Hu, A. Hamidi, S. Altemeyer, T. Koster, B. Spangenberg,
and H. Kurz, J. Vac. Sci. Technol. B {\bf 16}, 2822 (1998);
J.W. Lyding, T.C. Shen, J.S. Hubacek, J.R. Tucker, and G.C. Abein,
Appl. Phys. Lett. {\bf 64}, 2010 (1994); N. Suzuki, T. Makino,
Y. Yamada, T. Yoshida and  S. Onari, Appl. Phys. Lett. {\bf 76}, 1309
(2000); T. Toyama, Y. Kotani, A. Shimode, and H. Okamoto,
Appl. Phys. Lett. {\bf 73}, 105 (1998); E. Edelberg, S. Bergh,
R. Wame, M. Hall and E.S. Aydil, Appl. Phys. Lett. {\bf 68}, 1415
(1997); E.I. Givargizov, J. Vac. Sci. Technol. B {\bf 11}, 449 (1993);
A.M. Morales and C.M. Lieber, Science {\bf 279}, 208 (1998).

\bibitem{7} N. Miyata, H. Watanabe, and M. Ichikawa, Appl. Phys. Lett. 
{\bf 77}, 1620 (2000).

\bibitem{8} S. Facsko, T. Dekorsy, C. Koerdt, C. Trappe, H. Kurz, A. Vogt, and
H.L. Hartnagel, Science {\bf 285} (1999) 1551.

\bibitem{9} F. Frost, A. Schindler and F. Bigl, Phys. Rev. Lett., {\bf
85} (2000) 4116.

\bibitem{10} G. Carter and V. Vishnyakov, Phys. Rev. B, {\bf 54} (1996) 17647.

\bibitem{11} R. M. Bradley and J. M. E. Harper, J. Vac. Sci. Technol. A
{\bf 6}, 2390 (1988).

\bibitem{12} R. Cuerno and A.-L. Barab\'asi, Phys. Rev. Lett.
{\bf 74}, 4746 (1995).

\bibitem{13} F. Family and T. Vicsek, J. Phys. A {\bf18}, L15 (1985).

\bibitem{14} A.-L. Barab\'asi and H. E. Stanley, {\em Fractal Concepts in 
Surface Growth} (Cambridge University Press, Cambridge, 1995).

\bibitem{15} S. Park, B. Kahng, H. Jeong, and A.-L. Barab\'asi,
Phys. Rev. Lett. {\bf 83}, 3486 (1999). 

\bibitem{16} B. Kahng, H. Jeong, and A.-L. Barab\'asi, Appl. Phys. Lett. 
{\bf 78}, 805 (2001).

\bibitem{17} R. Gago, L. V\'azquez, R. Cuerno, M. Varela,
C. Ballesteros, and J. M. Albella, in preparation.

\end{thebibliography}
\end{document}